\documentclass[aps,prl,a4,twocolumn,superscriptaddress,nofootinbib,showpacs,showkeys,preprintnumbers]{revtex4-1}
\pdfoutput=1
\usepackage{color}
\usepackage[table]{xcolor}
\usepackage[colorlinks=true,citecolor=blue,linkcolor=blue,
breaklinks=true]{hyperref}
\usepackage{amsmath,amssymb}
\usepackage{graphicx}
\usepackage{slashed}
\usepackage[caption=false,labelformat=simple]{subfig}
\usepackage{wrapfig}
\usepackage{placeins}
\usepackage{multirow,makecell}
\usepackage{booktabs}
\usepackage{url}
\usepackage{enumitem}
\usepackage{multirow}
\usepackage{array}
\usepackage{tabulary}
\usepackage{float}

\newcommand{\lag}{\mathcal L}
\def\mET{\slashed{E}_T}

\begin{document}
\preprint{HRI-RECAPP-2024-04}
%%%%%%%%%%%%%%%%%%%%%%%%%  Title  %%%%%%%%%%%%%%%%%%%%%%%%%%
%\title{An emergent $Z^\prime$ from the Higgs shadow}
\title{Discovering an invisible $Z'$ at the muon collider}

%%%%%%%%%%%%%%%%%%%%%%%%%  Authors  %%%%%%%%%%%%%%%%%%%%%%%%

%\affiliation{Regional Centre for Accelerator-based Particle Physics, %\\ 
%Harish-Chandra Research Institute, HBNI, \\
%Chhatnag Road, Jhunsi, Prayagraj (Allahabad) 211\,019, India}

\author{Anjan Kumar Barik}
\email{anjanbarik@hri.res.in}
%\affiliation{Regional Centre for Accelerator-based Particle Physics, %\\ 
%Harish-Chandra Research Institute, HBNI, \\
%Chhatnag Road, Jhunsi, Prayagraj (Allahabad) 211\,019, India}

\author{Santosh Kumar Rai}
\email{skrai@hri.res.in}
\affiliation{Regional Centre for Accelerator-based Particle Physics, Harish-Chandra Research Institute,\\ 
A CI of Homi Bhabha National Institute, Chhatnag Road, Jhunsi, Prayagraj
211019, India.}

\author{Aviral Srivastava}
\email{aviralsrivastava12@gmail.com}
\affiliation{Delhi Technological University, \\ 
Shahbad Daulatpur, Main Bawana Road, Delhi 110042, India\\}

%%%%%%%%%%%%%%%%%%%%%%%%%  Abstract  %%%%%%%%%%%%%%%%%%%%%%%
\begin{abstract}
We show in this letter how a heavy $(\mathcal{O}(TeV))$ invisible $Z'$ gauge boson that will practically be out of reach of the Large Hadron Collider (LHC), can be 
discovered at the future muon collider. The new force carrier has a relatively stronger coupling with the beyond standard model (BSM) sector, while its interaction with the SM fields is much weaker. This weaker coupling is induced through mixing mechanisms, specifically via gauge kinetic mixing and the $Z-Z'$ mixing.
We consider a scenario where the new gauge boson decays mostly to charge-neutral long-lived particles and/or dark matter (DM). 
%The chances of observing such a heavy $Z'$ at the LHC is difficult due to its small coupling with SM particles.  Furthermore, the invisible nature of the $Z'$ makes its detection very challenging. 
We show how producing and detecting this heavier invisible $Z'$, that will be beyond the reach of even the very high luminosity LHC, becomes possible if it is produced in association with an energetic photon at the future muon collider. The on-shell production of the $Z'$ will lead to a peak in the photon energy distribution, following the so-called radiative return phenomena and can lead to the accurate determination of the $Z'$ mass and its interaction with SM particles.
\end{abstract}

\maketitle
%%%%%%%%%%%%%%%%%%%%%%%%%  Main Body  %%%%%%%%%%%%%%%%%%%%%%

%===========================================================
The Standard Model (SM) of particle physics, a theory of three fundamental interactions, successfully explains the physics below the subatomic level. Precision measurements of various experiments also show that there is not much deviation from the SM predictions. Although the SM is a renormalizable and unitary theory, it is highly unlikely that it is the ultimate theory of nature as it does not incorporate gravity. Many observed phenomena are unexplained by the SM e.g. neutrino mass \cite{Super-Kamiokande:1998kpq,LSND:2001aii,K2K:2002icj,T2K:2011ypd,DayaBay:2013yxg}, nature of the dark matter (DM) \cite{Zwicky:1933gu,Rubin:1980zd,Planck:2018vyg}, matter-anti-matter asymmetry \cite{Davidson:2008bu}, etc, which motivate us to look for the physics beyond SM (BSM).

There are a plethora of BSM models that have been proposed to explain various BSM phenomena. Models with $U(1)$ gauge extension to the SM symmetry are one of the simplest but very well-motivated BSM scenarios that have been studied in the literature \cite{Langacker:2008yv}. The motivation for an additional $U(1)$ gauge symmetry roots back to the Grand Unified Theories~(GUT), where GUTs with gauge symmetry of rank higher than four can give rise to additional $U(1)$ symmetry along with SM gauge group when the symmetry is spontaneously broken \cite{Langacker:1980js,Robinett:1981yz,Robinett:1982tq,Langacker:1984dc,Hewett:1988xc}. Additionally, the SM contains some global continuous symmetry which might be gauged, e.g. a non-anomalous linear combination of baryon and lepton numbers such as $B-L$, $L_\mu - L_\tau$, etc. \cite{Senjanovic:1975rk,Mohapatra:1980qe,FileviezPerez:2010gw}. Due to the presence of an extra $U(1)$ gauge symmetry that is broken spontaneously the model contains an additional heavy $Z'$ gauge boson. The $Z'$ can either couple to SM fields directly if the SM fields are charged under the new $U(1)$ gauge symmetry or it can couple to them via the gauge kinetic mixing (GKM) between the new $U(1)$ and SM hypercharge symmetry, $U(1)_Y$~\cite{Holdom:1985ag,Babu:1996vt,Babu:1997st,Foot:1991kb}. It has been shown that the GKM term is unavoidable at the one-loop level if there are particles present in a model that are simultaneously charged under the SM $U(1)_Y$ and new Abelian gauge symmetry unless there is a specific charge assignment \cite{delAguila:1988jz}. 
There is also a possibility that the new gauge boson can interact with the SM particles at one loop if there are particles that couple with both the SM particles and the $Z'$ \cite{Chauhan:2020mgv,Abdallah:2021dul,Bauer:2022nwt}. 

There have been searches for such additional gauge bosons in collider experiments for a wide range of mass windows, ranging from sub-MeV to a few TeV \cite{ALEPH:2013dgf,ATLAS:2019erb,LHCb:2017trq,ATLAS:2017eiz,ATLAS:2018mrn,CMS:2018hnz,ATLAS:2017ptz,CMS:2021fyk,CMS:2018yxg,ATLAS:2018coo,ATLAS:2019nat,CMS:2019qem,D0:2010kuq,BaBar:2014zli}. Null results from these experiments motivate us to consider a $Z'$ which might be either much heavier or couples to SM particles with tiny coupling strengths, that has helped it escape detection. A machine such as LHC with a total center-of-mass energy ($\sqrt{s}$) of the colliding beams at $14$ TeV should in principle be able to produce ultra-heavy particles on-shell. However, all the energy in the proton-proton collision is not available since the proton is a composite particle, and the hard scattering for any process depends on the constituent parton energies and their flux. Therefore producing heavy particles with masses on the order of $\mathcal{O}(\rm{TeV})$ on-shell results in a smaller cross section, as the parton flux goes down when the parton momentum fraction increases at the LHC. The LHC further distinguishes between new exotic particles based on whether they carry a colour quantum number or not. The lepton colliders that collide the elementary and stable particles are also sensitive to heavy particles through indirect exchanges or can produce them on-shell in association with another particle. The total energy of the beam is fixed in these collisions and unlike the LHC where an s-channel resonance of a heavy particle is possible due to the huge spread in $\sqrt{s}$, the lepton colliders will have to be tuned to the mass of the heavy particle to excite an s-channel resonance. Note that the electron-positron colliders have significant limitations in achieving collisions with very large $\sqrt{s}$. The electrons being very light, lose most of their energies in circular colliders via synchrotron radiation while at linear colliders they radiate its energy by the bremsstrahlung and beamsstrahlung processes which affects the beam shape at collision, thus affecting the luminosity of collision. In contrast, the muons being $\sim 200$ times heavier than electrons can be accelerated to very high energy at circular colliders due to smaller radiation \cite{Chen:1993dba,Barklow:2023iav}.  Therefore, the muon collider proposed by the International Muon Collider Collaboration (IMCC) is an important future machine that would help explore BSM scenarios at very high center-of-mass energies. This machine is proposed to run with $\sqrt{s}=3,10$, and $14$ TeV \cite{Han:2020uid,AlAli:2021let,Accettura:2023ked,MuonCollider:2022xlm,Delahaye:2019omf,Schulte:2022brl}. 
%Another strong motivation for the muon collider is to explore muphilic interactions \cite{Lu:2023ryd,Han:2021lnp,Dermisek:2021mhi,Celada:2023oji,Dermisek:2023rvv,Aime:2022flm,Black:2022cth,Ghosh:2022vpb,Ghosh:2023xbj} guided by the $\sim 5 \sigma$ deviation of muon $g-2$ from the SM prediction \cite{Muong-2:2023cdq,Venanzoni:2023mbe}. 
%To study BSM physics we should consider a process where a BSM particle can be produced on-shell the reason is that the off-shell particle production suffers a propagator suppression therefore leading to a smaller cross-section. 
%In the muon collider, the initial state radiations (ISR) take a very small fraction of beam energy, therefore, the cross-section of producing particles on-shell in an s-channel process with mass $M_s<\sqrt{s}$ would be very small, where $\sqrt{s}$ is the centre mass-energy. Whereas, these particles can be produced associated with other particles. 

We must note that any resonant production of a BSM particle ($X$) of mass $M_X$ at the muon colliders (in the $s$-channel) will be possible if the $\sqrt{s}\simeq M_X$. So if the particle mass $M_X \ll \sqrt{s}$, the production cross-section will be propagator suppressed. However, if the particle is produced in association with another particle which carries away a significant amount of energy, we can produce $X$ on-shell and study its properties. One such channel is an associated photon signal i.e. $\ell^- \ell^+ \to \gamma X $, where $\ell$ corresponds to the colliding leptons. In this simple $2 \to 2$ process, the photon energy depends only on the center-of-mass energy and the mass of $X$ can be determined by using $E_\gamma = \frac{s-M^2_X}{2\sqrt{s}}$. In a realistic scenario this will demonstrate a peak in the observed $E_{\gamma}$ distribution. As the $\sqrt{s}$  of the machine is fixed, this peak will give a clear information about the mass of the particle $X$, even if it is long-lived or decays invisibly. The associated photon production along with SM weak gauge bosons was used by the Large Electron Positron (LEP) experiment to measure the gauge interactions and the anomalous coupling of the photon with SM weak gauge bosons \cite{Denner:2001vr,L3:2001yhb}. Similarly the measurement of the cross-section of the process $e^- e^+ \to \gamma\nu \bar{\nu}$ at LEP was able to determine the precise number of light neutrino generations in the SM \cite{Ma:1978zm}. Therefore if there exists any hidden particle in nature that couples to SM leptons, the process $\ell^- \ell^+ \to \gamma \mET$ can give a hint about the hidden particle as this could leave an imprint about that particle mass in the
photon energy distribution \cite{Chen:1974wv,Fayet:1982ky,Ellis:1982zz,Grifols:1985ix,Datta:1994ac,Chen:1995yu,Datta:2002jh,KumarRai:2003kk,Choudhury:2004ea,Dreiner:2006sb,Rai:2008ei,Chakrabarty:2014pja}. The same feature can be exploited at the muon collider to search for particles that are very weakly coupled to the SM and their masses are beyond the reach of LHC. Note that with the photon being massless, the mass reach for the new particle can extend from being much less than $\sqrt{s}$ to $\sqrt{s}$ itself. In this letter, we consider this channel to probe the presence of a hidden new gauge boson $Z'$ which couples to the BSM sector maximally, at the muon collider. We note that this can be generalized to study different heavy states.
%, we consider a  $U(1)$ extension of SM where no SM particles are charged under the new gauge group and can only interact with the new gauge boson $Z'$ due to the presence of the GKM term between the new $U(1)$ symmetry and the SM $U(1)_Y$ symmetry. Due to the presence of three generations of vector-like neutral fermions, the SM neutrinos become massive via the inverse seesaw mechanism.   

We consider a model that is an extension of the SM where the gauge symmetry is extended by an additional $U(1)$ symmetry and is responsible for neutrino mass generation via the inverse-seesaw mechanism \cite{Abdallah:2021npg}. 
%Besides the SM field content, the model contains two additional scalars $H_2$ and $S$ and three generations of left and right-handed Weyl fermions. The presence of $H_2$ ensures the mass for the SM neutrinos via the inverse-seesaw mechanism when it gets vacuum expectation value (vev). Both $H_2$ and $S$ are responsible for the breaking of new $U(1)$ symmetry spontaneously when they acquire vev. 
The charge assignment of the relevant BSM fields along with the SM like Higgs doublet $H_1$ under the gauge symmetry of the model is shown in Table \ref{tab:charges}. 
\begin{table}[h!]
\begin{center}
\begin{tabular}{|c|c|c|c|c|c|}
\hline %& & & & &\\[-3mm]
Fields  & $SU(3)_C$ & $SU(2)_L$ & $U(1)_Y$ & $U(1)_X$ & Spin \\[1mm]
\hline %& & & & &\\[-3mm]
$H_{1(2)}$  & 1 & 2 & $-1/2$ & 0 ($-\, 1$) & 0 \\ [1mm]
\hline %& & & & &\\[-3mm]
$S$ & 1 & 1 & 0 & $2$ &  0 \\ [1mm]
\hline %& & & & &\\[-3mm]
$N_{L/R}^i$ & 1 & 1 & 0 & $1$  & 1/2 \\ [1mm]
\hline %& & & & &\\[-3mm]
\end{tabular}
\end{center}
\caption{Scalars ($H_1,H_2, S$) and BSM fermions ($N_L^i, N_R^i, \, i=1,2,3$) and their charge assignments under the SM gauge group and $U(1)_X$.}
\label{tab:charges}
\end{table}
%===========================================================
%%%%%%%%%%%%%%%%%%%%%%%%%%%%%%%%%%%%%%%%%%%%%%%%%%%%%%%%%%%%
%Both $H_2$ and $S$ are responsible for the breaking of new $U(1)$ symmetry spontaneously when they acquire vev.
%The Lagrangian of the BSM fields and $H_1$ is given by (ignoring the kinetic term)
%\begin{eqnarray}
%&\lag_S& \supset  - \mu_1 H_1^\dagger H_1\! -\! \mu_2 H_2^\dagger H_2 \!-\! \mu_s S^\dagger S  
 %                                     \!+\! \{\mu_{12} H_1^\dagger H_2 + {\rm\, h.c.} \}  \nonumber \\
 %&-&   \lambda_1 (H_1^\dagger H_1)^2 - \lambda_2 (H_2^\dagger H_2)^2  
 %            - \lambda_s (S^\dagger S)^2  - \lambda'_{12} \left|H_1^\dagger H_2\right|^2  \nonumber \\
 % &-&   \lambda_{12} H_1^\dagger H_1 H_2^\dagger H_2  - \lambda_{1s} H_1^\dagger H_1 S^\dagger S 
% - \lambda_{2s} H_2^\dagger H_2 S^\dagger S,  \nonumber \\[0.1cm] 
%&\lag_{Y}& \supset  - \{ Y_\nu\,\overline l_L H_2 N_R \!+\! Y_R S \overline N_R N_R^C \!+\! Y_L S \overline N_L N_L^C \!+\! {\rm\, h.c.}\},  \nonumber \\[0.1cm] 
%&\lag_{M}& \supset  - \hat{M}_N\left(\overline{N}_L N_R + \overline{N}_R N_L \right). 
%\label{eqn:lag}
%\end{eqnarray}
%Note that the term with the coefficient $\mu_{12}$ explicitly breaks the $U(1)_X$ gauge symmetry. Without this term, the model contains a real goldstone boson due to the presence of a global $U(1)$ symmetry in the scalar potential which is spontaneously broken when the scalars acquire vev. The term can be generated in a UV complete scenario by adding another singlet scalar with $U(1)_{X}$ charge opposite to the $H_2$. 
The Lagrangian containing the kinetic term of the scalars and the fermions along with their interaction with gauge bosons is given by 
\begin{eqnarray}
\lag &\supset& \left(D_\mu H_1 \right)^\dagger D_\mu H_1 + \left(D_\mu H_2 \right)^\dagger D_\mu H_2 + \left(D_\mu S \right)^\dagger D_\mu S  \nonumber \\
&       & +\ i\,\overline N_L \gamma^\mu D_\mu N_L + i\,\overline N_R \gamma^\mu D_\mu N_R .
\label{eqn:lag}
\end{eqnarray}
The covariant derivatives for the fields $H_1,H_2,S$ and $N_{L/R}$ are defined as
\begin{align}
D_\mu^{(H_1)} &= \partial_\mu -ig_2 \frac{\sigma^a}{2}W_\mu^a + i\frac{g_1}{2}B_\mu  \, ,\nonumber \\
D_\mu^{(H_2)} &= \partial_\mu -ig_2\frac{\sigma^a}{2}W_\mu^a +i\frac{g_1}{2}B_\mu + i g_x C_\mu  \, , \nonumber \\
D_\mu^{(S)} &= \partial_\mu -2ig_x  C_\mu  \, , \,\,\,\, %,\nonumber \\
D_\mu^{(N_{L/R})} = \partial_\mu -ig_x C_\mu  \, , 
\end{align}
where  $B_\mu$, $W_\mu^a$, and $C_\mu$ are the gauge fields of the $SU(2)_L, U(1)_Y\, \text{and} \, U(1)_X$ gauge symmetry respectively with their corresponding gauge couplings given by $g_1$, $g_2$, and $g_x$.  
The spontaneous breaking of electroweak and $U(1)_{X}$ gauge symmetry occurs when the scalars $H_1$,$H_2$ and $S$ acquire vev $v_1$,$v_2$ and $v_s$ respectively. It is quite clear that $C_\mu$ only couples to the new fields ($N, S, H_2$) introduced in the model. There are three CP even scalars present in the model whose mass eigenstates are denoted as $h_1$, $h_2$, and $h_s$.  The $h_1$, which dominantly comes from the doublet $H_1$ is identified as the SM Higgs. 
%this can be achievable by choosing small $\tan\beta =  \frac{v_2}{v_1}$  and the smaller quartic coupling of $H_1$ with other scalars. 
The model also contains one physical CP odd scalar $(A)$ and one charged scalar $(H^\pm)$. As our focus in this Letter is solely on the new gauge boson, we refer our readers to 
Refs.~\cite{Abdallah:2021dul,Abdallah:2021npg} for more details on the model's particle spectrum and its parameters.

The Lagrangian containing all the gauge fields in the model is given by 
\begin{eqnarray}
	\lag \supset &-& \frac{1}{4} G^{a,\mu\nu} G_{\mu\nu}^a - \frac{1}{4} W^{b,\mu\nu} W_{\mu\nu}^b -\frac{1}{4} B^{\mu\nu} B_{\mu\nu}  \nonumber \\ 
    &-& \frac{1}{4} C^{\mu\nu} C_{\mu\nu} + \frac{1}{2} \tilde{g} B^{\mu\nu} C_{\mu\nu} \, , 
\end{eqnarray}
 where $G^{a}_{\mu\nu}, W^{b}_{\mu\nu}, B_{\mu\nu}\, \rm{and} \, C_{\mu\nu}$ are the field strength tensor of $SU(3)_C, SU(2)_L, U(1)_Y\, \text{and} \, U(1)_X$ gauge bosons respectively. As the field strength tensor of a $U(1)$ gauge field is gauge invariant, therefore, a GKM term is allowed between  $U(1)_Y\, \text{and} \, U(1)_X$ gauge bosons in this model, which is governed by the $\tilde{g}$ term in the above Lagrangian. 
 %The electroweak symmetry is spontaneously broken to $U(1)_{em}$ when $H_1$ and $H_2$ get vev and similarly the $U(1)_X$ symmetry is broken when $H_2$ and $S$ get vev. As the model has additional gauge symmetry, three neutral colourless gauge bosons exist in this model. 
 After symmetry breaking, and rotating the fields to their mass eigenbasis, we get the photon, the SM $Z$ boson and a new gauge boson $Z'$ as the mediators of gauge interactions. The masses and mixing angle of the $Z$ and $Z'$ are given by 
{\small
\begin{eqnarray}
M_{Z,Z'}^2 &=& \frac{1}{8}\Big[ (g_z^2  + {g'_x}^2) v^2 + 4 C v_2^2\Big] \nonumber\\
& &  \mp   \frac{1}{8}\sqrt{\Big({g'_x}^2 v^2 + 4 C v_2^2 - g_z^2 v^2\Big)^2 + 4 g_z^2 D^2}  \nonumber \, ,\\ 
\tan2\theta' &=& \dfrac{2g_z \left(g'_x v^2 + 2g_x v_2^2\right)}{{g'_x}^2 v^2 + 4C v_2^2  
- g_z^2 v^2} \,,
\end{eqnarray} }
where $g_z=\sqrt{g_1^2+g_2^2}, \, g'_x=\frac{g_1\tilde{g}}{\sqrt{1-\tilde{g}^2}}, \, C= g_x g'_x  +  g_x^2(1+ 4 \frac{v_s^2}{v_2^2}), \, D = \Big(g'_x v^2 + 2 g_x v_2^2\Big)$ and $v=\sqrt{v_1^2+v_2^2} \simeq 246 \,{\rm GeV}$.

The new gauge boson $Z'$ couples to SM fermions via the $Z-Z'$ mixing angle $\theta'$ and the redefined GKM parameter $g'_x$. 
%Its coupling with SM and the BSM fermions has been given in the appendix. 
%The mixing angle $\theta'$ can not be very large as it would alter the well-measured $Z$ boson decay width and branching fraction to SM fermions. 
The $Z-Z'$ mixing angle is constrained to values $\leq 10^{-3}$ by the
LEP~\cite{ParticleDataGroup:2020ssz}.    

%The SM neutrinos become massive via the inverse seesaw mechanism. 
The Yukawa part of the Lagrangian which couples the scalars and the new SM singlet vector-like fermions is
\begin{align}
\lag_{Y} \supset  - \{ Y_\nu\,\overline l_L H_2 N_R \!+\! Y_R S \overline N_R N_R^C \!+\! Y_L S \overline N_L N_L^C \!+\! {\rm\, h.c.}\}, % \nonumber \\[0.1cm] 
%&\lag_{M}& \supset  - \hat{M}_N\left(\overline{N}_L N_R + \overline{N}_R N_L \right). 
\label{eqn:lag}
\end{align}
where the singlet fermions have a mass term $\lag_{M} \supset  - \hat{M}_N\left(\overline{N}_L N_R + \overline{N}_R N_L \right)$.
The SM neutrinos mix with the BSM fermions due to the presence of Yukawa coupling $Y_\nu$. 
%For simplicity, we take $M_N$ to be diagonal. 
The model has six heavy Majorana neutrino states $(\nu_{4-9})$ in addition to the SM neutrinos which get their mass via inverse seesaw mechanism.
%After diagonalizing the mass matrix of the neutral fermions we obtain nine Majorana fermions three lightest among them are SM neutrinos while six other $(\nu_{4-9})$ are named as heavy neutrinos. 
An interesting feature of the model is that by appropriately choosing the Yukawa coupling ($Y_\nu$) in the $Y_\nu\,\overline l_L H_2 N_R$ term, we can make the heavy neutrino very long-lived and for $Y_\nu \sim 0$ it can also become a DM candidate~\cite{Abdallah:2024npf}. 
%Thus, a decaying DM scenario arises in this model, where the DM can decay into SM leptons and gauge bosons. The most stringent bound for heavy neutrino-type DM comes from the IceCube collaboration, which suggests that the Yukawa coupling must be  $\leq \mathcal{O}(10^{-27})$ \cite{Higaki:2014dwa,BhupalDev:2016gna,ReFiorentin:2016rzn,DiBari:2016guw}. In this letter, we choose $Y_{\nu_{11}}=10^{-27}$, $Y_{\nu_{1i}}=0$, therefore, two heavy neutrinos $\nu_{4}$ and $\nu_5$ become long-lived.
%Assuming that the lightest of them $(\nu_{4})$ becomes the DM candidate. The $\nu_5$ decays to $\nu_4$ and an SM fermion anti-fermion pair via the off-shell $Z$ and $Z'$. The DM phenomenology of this model has been explored in a recent work \cite{Abdallah:2024npf}.

%The searches for the $Z'$ resonance have been carried out at the LHC in various final states such as the dilepton, $t\bar{t}$, $b\bar{b}$ and $jj$, null results from these experiments put bounds on its coupling with these SM fermions. Other $Z'$ resonance search channels involve its decay to $Z h$ with $Z\to \ell^- \ell^+, \nu \bar{\nu}$ and $h \to b\bar{b}$ has been explored by ATLAS and CMS collaborations. There also exists $Z'$ searches with associated production at LHC i.e. $p p \to Z' X \to W^+ W^- X$, where $W$ boson decay hadronically. An interesting search for light $Z'$ has been looked at the Higgs resonance channel followed by its decay to a pair of $Z'$ then the $Z'$ subsequently decays to leptons resulting in four lepton final states studied by the ATLAS collaboration. 
In most searches of the $Z'$ it is assumed to have direct couplings to the SM fermions. The $Z'$ is then assumed to decay to SM particles with $100\%$ branching fraction~\cite{ALEPH:2013dgf,ATLAS:2019erb,LHCb:2017trq,ATLAS:2017eiz,ATLAS:2018mrn,CMS:2018hnz,ATLAS:2017ptz,CMS:2021fyk,CMS:2018yxg,ATLAS:2018coo,ATLAS:2019nat,CMS:2019qem,D0:2010kuq,BaBar:2014zli}. These searches put a stringent constraint on its mass and couplings to SM particles as no $Z'$  signal has been observed. These bounds can be relaxed significantly if one considers a scenario where $Z'$ decays mostly to BSM particles~\cite{Abdallah:2021npg}. 
%We have also studied the case where the $Z'$ couples weakly to SM fermions, the standard s-channel $Z'$ resonance from the fermions would lead to a smaller cross-section which makes its detection challenging therefore we rely on the Higgs resonance for its pair production and explored the four lepton final state. 
%%%%%%%%%%%%%%%%%%%%%%%%%%%%%%%%%%%%%%%%%%%%%%%%%%%%%%
\begin{figure}[t!]
\begin{center}
\includegraphics[width=7cm,height=6cm]{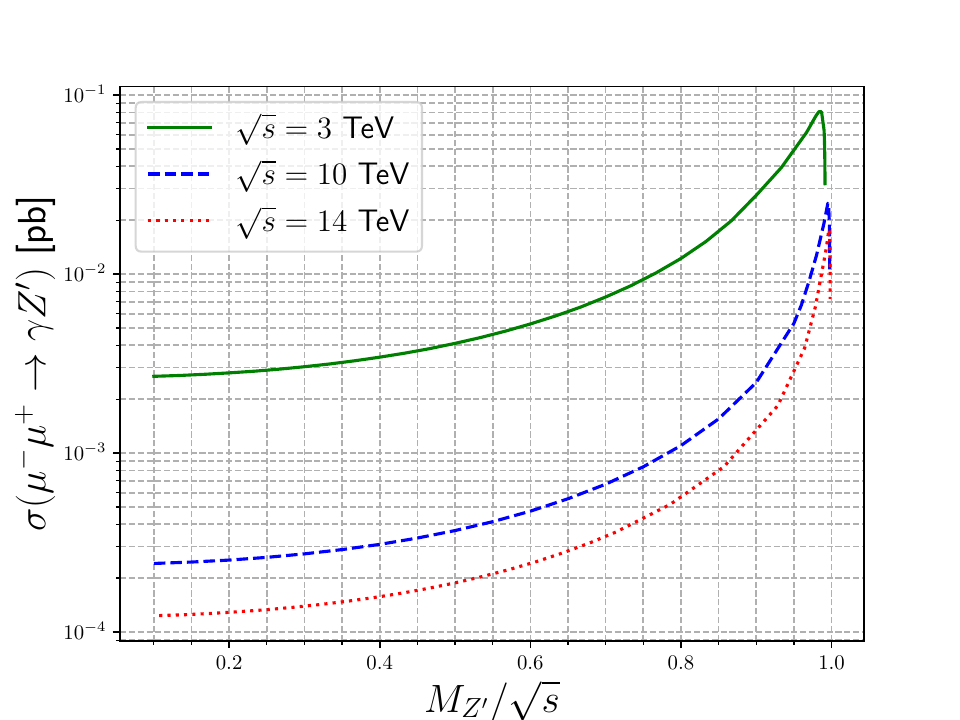}
\end{center}
\vspace{-0.5cm}
\caption{Variation of the cross section for the process $\mu^- \mu^+ \to \gamma Z'$ as a function of $\frac{M_{Z'}}{\sqrt{s}}$ for three center-of-mass energies with $g'_x = 0.05$ and $p_{T_{\gamma}}\geq 25$ GeV. }
\label{Fig-crx}
\end{figure}
%%%%%%%%%%%%%%%%%%%%%%%%%%%%%%%%%%%%%%%%%%%%%%%%%%%%%
In this letter, we explore the scenario in which the $Z'$ predominantly decays into $\nu_{4}$ and $\nu_5$ that may be very long-lived or be stable, which implies an invisible decay of the $Z'$ in the detector. We consider the associated production of $Z'$ along with the photon at the muon collider, {\it viz.} $\mu\mu \to \gamma Z'$ followed by the decay of $Z' \to \nu_{4} \nu_{5} $. Note that the production of the $Z'$ at the muon collider depends on the GKM parameter $g'_x$, while its decay to $\nu_{4} \nu_{5} $ depends on the gauge coupling $g_x$. The $Z'$ will decay dominantly to the invisible BSM mode when 
%%%%%%%%%%%%%%%%%%%%%%%%%%%%%%%%%%%%%%%%%%%%%%%%%%%%%%%%%%%%%%%%%%%%%%%%%%%%%%%%%%%%%%%%%%%%%%%%%%%%%%%%%%
\begin{table}[b!]
%\centering
\begin{center}\scalebox{0.90}{
\begin{tabular}{|c|c|c|c|c|c|}
\hline &  $M_{Z'}$ (TeV) &  $M_{\nu_4} $ (GeV) & $g_x ' $ & $\tan\theta' \times10^{4}$ & BR $(Z' \to \nu_{4} \, \nu_{5})$ \\ \hline
%\hline
     {\tt BP1} & 1.0      & 407.92        &0.04  & 4.5995 &0.987           \\
     {\tt BP2} & 2.0      & 856.92        &0.057  & 1.6280 &  0.971        \\ 
     {\tt BP3}   &  2.9          &1299.38   & 0.05     &0.6785 &0.975        \\ 
      {\tt BP4}   &  5.0          & 2310.88   & 0.05     &0.2281 & 0.972       \\ 
 {\tt BP5} & 8.0     &  3771.88      & 0.05   &0.0891 &  0.968     \\
 {\tt BP6}  & 9.8    & 4655.11    & 0.05  & 0.0593 & 0.966    \\     
\hline
 \end{tabular}}
\end{center}
\caption{Benchmark points showing the $Z'$ and DM masses, the corresponding GKM parameter $g_x'$, 
$Z-Z'$ mixing angle ($\theta'$), and the invisible branching ratio of $Z'$ decaying to $\nu_{4} \, \nu_{5}$ mode. Note that we have kept $\tan\beta = v_2/v_1 \simeq 10^{-4}$ and $g_x = 0.9$ for all the 
{\tt BP}s. All other particles that are charged under $U(1)_X$ are assumed to be heavier than $Z'$. 
%All these {\tt BP}s satisfy DM relic density $\Omega h^2=0.1198\pm 0.0012$ within $3\sigma$ and also are allowed by the direct detection experimental data.
}
\label{tab:bps}
\end{table}
%%%%%%%%%%%%%%
$g_x >> g'_x$. As the production of $Z'$ in the above process is through the GKM parameter $g'_x$, it should not be very small. The dilepton resonance search at the LHC constraints the GKM parameter \cite{ATLAS:2019erb} while experimental constraints on the effective heavy scale for four-fermion contact interactions \cite{ParticleDataGroup:2020ssz} will also put a bound on $g'_x$. The production cross-section of the process $\mu^- \mu^+ \to \gamma\, Z'$ as a function of $\frac{M_{Z'}}{\sqrt{s}}$ for three center-of-mass energies $(3, 10, 14)~\rm{TeV}$ at the future muon collider has been shown in the Fig.~\ref{Fig-crx}. The GKM parameter $g'_x$ is taken to be $0.05$ for this plot with the minimum transverse momentum $(p_{T})$ of the photon to be $25$ GeV. From the figure, we find that the cross-section of this process increases as we go to the higher values of $M_{Z'}$ for a given $\sqrt{s}$. As the $M_{Z'}$ approaches $\sqrt{s}$ of the machine, the $Z'$ can be produced as a $s$-channel resonance ($\mu\mu \to Z' (\gamma)$), leading to significant increase in the cross-section. The fall in the cross-section as $\frac{M_{Z'}}{\sqrt{s}}\sim 1$  is due to the phase space suppression on account of the photon carrying a minimum $p_{T_{\gamma}}\geq 25$ GeV.

Before going to the details of the analysis we would like to mention that this model has been implemented in {\tt SARAH} \cite{Staub:2015kfa} to obtain the {\tt Universal Feynman Object (UFO)}. Then these {\tt UFO} files have been used to generate the events for muon collider in {\tt MadGraph5@aMCNLO} \cite{Alwall:2014hca} using the spectrum file which is generated by the package {\tt SPheno} \cite{Porod:2011nf}. The events are then passed through {\tt Pythia 8} \cite{Sjostrand:2014zea} for the detector simulation using {\tt Delphes-3} \cite{deFavereau:2013fsa} using the muon detector card. The cut-based analysis is done using the package {\tt MadAnalysis5} \cite{Conte:2012fm}. We have chosen the lightest heavy neutrino to be the DM candidate in our analysis and the DM relic density calculation has been done using package {\tt micrOMEGAs}\cite{Belanger:2010pz}. 

%%%%%%%%%%%%%%%%%%%%%%%%%%%%%%%%%%%%%%%%%%%%%%%%%%%%%%%%%%%%%%%%%%%%%%%%%%%%%%%%%%%%%%%%%%%%%%%%%%%%%%%%%
\begin{table}[t!]
%\centering
\begin{center}\scalebox{1.0}{
\begin{tabular}{|c|c|c|}
\hline &  $\sigma (\sqrt{s}=3$ TeV) (fb)&  $\sigma (\sqrt{s}=10$ TeV )(fb)\\ \hline
%\hline
     SM & 2118.01      & 2452.02                   \\
     {\tt BP1} &  2.08     &  0.161                  \\
     {\tt BP2} & 8.75      &  0.339             \\ 
     {\tt BP3}   &  48.94         & 0.281       \\ 
      {\tt BP4}   & $2.436\times 10^{-5}$          & 0.387     \\ 
 {\tt BP5} & $2.295\times 10^{-6}$     &  1.16         \\
 {\tt BP6}  &  $9.396\times 10^{-7}$   &   9.45      \\     
\hline
 \end{tabular}}
\end{center}
\caption{Cross section for the for the process $\mu^- \mu^+ \to \gamma \,\nu_{i}\nu_{j} (i,j = 1,2,3~(\rm{SM})\, \rm{or}\, 4,5~(\rm{BSM}))$.}
\label{tab:crx}
\end{table}
%%%%%%%%%%%%%%%%%%%%%%%%%%%%%%%%%%%%%%%%%%%%%%%%%%%%%%%%%%%%%%%%%%%%%%%%%%%%%%%%%%%%%%%%%%%%%%%%%%%%%%%%%%%%
\begin{figure}[h!]
\begin{center}
{\includegraphics[width=6cm,height=6cm]{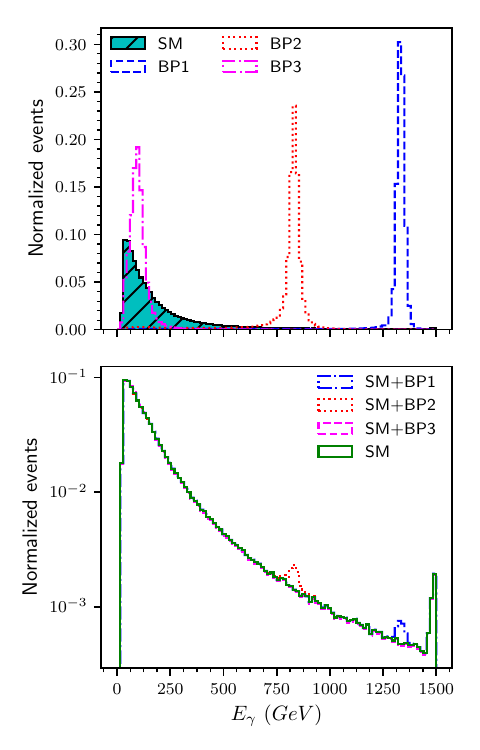}\label{Fig-3TeV-a}}
{\includegraphics[width=6cm,height=5cm]{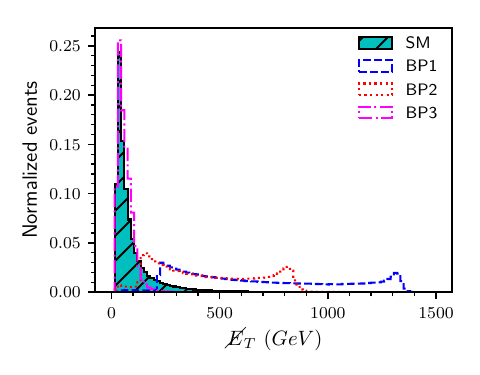}\label{Fig-3TeV-b}}
\end{center}
\vspace{-0.5cm}
\caption{Normalized event distributions of photon energy for signal and SM (top), combined SM+signal (middle) and missing traverse energy, $\mET$ (bottom), at muon collider with $\sqrt{s} = 3~\rm{TeV}$ for the BPs $\tt{BP1, BP2}$ and $\tt{BP3}$.}
\label{Fig-3TeV}
\end{figure}
%%%%%%%%%%%%%%%%%%%%%%%%%%%%%%%%%%%%%%%%%%%%%%%%%%%%%%%%%%%%%%%%%%%%%%%%%
\begin{figure}[h!]
\begin{center}
{\includegraphics[width=6cm,height=5cm]{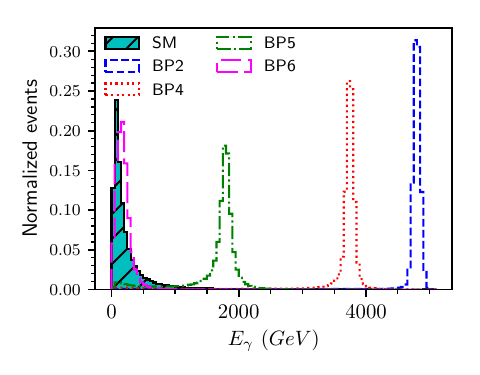}\label{Fig-10TeV-a}}
\end{center}
\vspace{-0.5cm}
\caption{Normalized event distribution of photon energy for signal and SM at muon collider with $\sqrt{s} = 10~\rm{TeV}$ for the BPs $\tt{BP2, BP4, BP5}$ and $\tt{BP6}$. }
\label{Fig-10TeV}
\end{figure}
%%%%%%%%%%%%%%%%%%%%%%%%%%%%%%%%%%%%%%%%%%%%%%%%%%%%%%%%%%%%%%%%%%%%%%%%%%%%%%%%%%%%%%%%%%%%%%%%%%%%%%%%%%%%
We study the $\mu^- \mu^+ \to \gamma + \mET$ channel for $\sqrt{s}=3, 10$ TeV. The SM background for the signal comes from $\mu^- \mu^+ \to \gamma \nu \bar{\nu}$ process whereas the source of the missing energy in the BSM signal is due to the decay of $Z' \to \nu_4 \nu_5$, where the DM $(\nu_{4})$ and the long-lived particle $(\nu_{5})$ in the final state 
pass through the detector undetected.  We choose six representative Benchmark Points ($\tt{BP}$) for this analysis that are shown in Table \ref{tab:bps}. 
The $\nu_4$ and $\nu_5$ masses are chosen to resonantly co-annihilate into SM particles via the 
$s$-channel $Z'$ process that determines the DM relic density~\cite{Abdallah:2024npf}. 
The mass of the DM has been shown for each {\tt BP} in Table \ref{tab:bps} and the mass of the $\nu_5$ lies within $M_{\rm{DM}}+\mathcal{O}(1-100)~\rm{MeV}$. The $\nu_5$ has a 
decay width of $\mathcal{O}(10^{-22}-10^{-23})$ GeV, making it very long-lived. 
%The dominant annihilation channels for obtaining correct DM relic density are 
%$\nu_4 \nu_5 \to f \bar{f}$ via $Z'$, where $f$ and $\bar{f}$ are SM fermion and anti fermion. The $\nu_4$ and $\nu_5$ annihilating to $W^+ W^-$ and $h_1 Z$ are subdominant as they depend on $\sin{\theta'}$ in contrast to the other processes which rely on the GKM parameter. 
%For the considered parameter region the co-annihilation process via $Z'$ determines the DM relic density as for the small mass difference between $\nu_4$ and $\nu_5$, $\nu_4 \nu_4 Z'\, (\nu_5 \nu_5 Z')$ becomes negligible. 
All the chosen {\tt BP}'s %We have also checked that 
give a DM relic density within $3\sigma$ of $\Omega h^2=0.1198\pm 0.0012$ and they all satisfy the constraints from direct and indirect search experiments~\cite{Abdallah:2024npf}. %allow these {\tt BP}s. The DM nucleon spin-dependent scattering cross section for these {\tt BP}s are coming out to be $\leq \mathcal{O}(10^{-19})$ pb and the average annihilation cross section times velocity $(\langle \sigma v \rangle)$ of the DM is $\leq \mathcal{O}(10^{-37})\, cm^3/s$. Therefore probing this DM in direct and indirect experiments would be very challenging. But due to the presence of the mediator $Z'$ which couples to both the $\nu_4$, $\nu_5$, and SM, thus can be produced in the collider experiments. 

We have generated events for the full $2 \to 3$ process, $\mu^- \mu^+ \to \gamma + \nu_4 + \nu_5$ which gives the dominant contribution as well as $\mu^- \mu^+ \to \gamma + \nu_i + \nu_j$ ($i,j=1,2,3$) that gives negligible contributions, where both subprocesses are mediated by the $Z'$.  
For the analysis of our signal we have selected  only events with a single hard photon in the final state that has $p_T > 25$ GeV and $|\eta_\gamma| < 3.0$. The signal and background cross-section with this selection cut are shown in Table \ref{tab:crx}. Note that for $\sqrt{s}=3$ TeV, the signal has large cross sections for 
$\tt{BP1, BP2, BP3}$ as long as 
$(M_{\nu_4}+M_{\nu_5}) < \sqrt{s}$, and the 
on-shell production of the $Z'$ is possible. The cross sections drop to very small values for the remaining benchmark points as the signal only gets contributions from suppressed couplings of $Z'$ to SM neutrinos ($\mu \mu \to \gamma+(Z'^* \to \nu_i \nu_j) \, i,j=1,2,3$). However, for 
$M_{Z'} = 5~{\rm TeV} > \sqrt{s}$ but with $(M_{\nu_4}+M_{\nu_5}) < \sqrt{s}$ the signal cross section is only suppressed at the 
$Z'\mu\mu$ vertex and we get a slightly bigger signal cross section of $\sim 2\times 10^{-2}$ fb. We restrict our analysis to cases where $M_{Z'} < \sqrt{s}$ where we expect to observe the "radiative return" peak in the photon energy $(E_{\gamma})$ distribution.
%The signal-to-background ratio can be increased by observing the various kinematic observables and choosing appropriate cuts on these variables to reduce the background. 
The normalized distribution of signal and background events as a function of $E_{\gamma}$ and missing transverse momentum $(\mET)$ are shown in 
Fig.~\ref{Fig-3TeV} at $\sqrt{s}=3$ TeV. In Fig.~\ref{Fig-10TeV} the photon energy distribution is shown for $\sqrt{s}=10$ TeV. As expected, for the SM background the neutrinos are produced back-to-back and the differential cross section peaks when the photon is soft. Here the dominant contribution comes from the $t$-channel exchange of the $W$ boson and the $Z$ mediated $s$-channel is heavily propagator suppressed. The differential cross section falls smoothly as we go to higher photon energy and the $t$-channel process gets more and more suppressed until 
$E_\gamma = \frac{s-M_Z^2}{2\sqrt{s}}$ where the $Z$ exchange dominates the SM differential cross section at this high $E_\gamma$ bin. As seen in the middle plot of Fig.~\ref{Fig-3TeV}, the SM background hits the radiative return peak of $Z$ boson around the bin where $E_\gamma \simeq 1498$ GeV. 
%most of the total COM energy is taken by neutrino anti-neutrino pair resulting in photon energy peaking near low energy.  This can be understood as follows neutrino and antineutrino pairs can be produced from the t-channel exchange of $W$ boson ($Z$ exchange is sub-dominant as COM energy is much away from its resonance) and the photon can be radiated from a real and virtual charged particle to balance the energy-momentum. The phase space distribution favors the invariant mass of photon-neutrino and photon-antineutrino system peaks at smaller energy,  as this process has a symmetry in exchanging particles with their antiparticles neutrino and antineutrino have identical distribution in all the kinematic variables they contain significant of available energy leading the photon energy to peak at low energy. 
In the BSM scenario where $Z'$ is produced on-shell along with photon and then decays to $\nu_4\, \nu_5 $, one expects a similar phenomena to what was seen as the $Z$ peak in the SM background distribution. In this case the photon energy peaks at  $E_{\rm{peak}} = \frac{s-M^2_{Z'}}{2\sqrt{s}}$. This peak is very sharp and will gradually shift towards the soft part of the photon distribution for heavier $Z'$. Using the position of the peaks in the distribution that sit above the SM background for all the {\tt BP}s, appropriate selection conditions on the events in $E_\gamma$ can help discover the $Z'$ efficiently 
%%%%%%%
\begin{table}[t!]
\centering
\resizebox{8cm}{!}{
\begin{tabular}{|c|c|c|c|c|c|c|}
\cline{2-7}
\multicolumn{1}{c|}{$\mathcal{L}=100$ fb$^{-1}$} & \multicolumn{6}{c|}{$\gamma+\mET$} \\  \hline
                          & \multicolumn{2}{c}{} & \multicolumn{2}{c}{Events} &  \multicolumn{2}{c|}{} \\ \hline
                       Cuts & SM & \tt{BP1} & SM & \tt{BP2} & SM &\tt{BP3} \\ \hline
        $N_\gamma=1$ & $167618$  & $154$ & $167618$ & $663$ & $167618$ & $4421$ \\ \hline
$E_{\gamma} \in E_{\rm{peak}}\mp 200$ GeV & $2982$ & $146$ & $6917$ & $606$ & $134938$ &  $4387$\\  \hline 
$E_{\gamma} \in E_{\rm{peak}}\mp 100$ GeV & $1124$ & $145$ & $3290$ & $574$ & $114504$ &  $4283$\\  \hline \hline
    \multicolumn{1}{|c|}{Significance ($\mathcal{S}$)} & \multicolumn{2}{c|}{$4.23$} & \multicolumn{2}{c|}{9.73} &  \multicolumn{2}{c|}{12.58} \\ \hline
	\end{tabular}}
 \caption{\small The cutflow information of SM background, {\tt BP1}, {\tt BP2} and 
 {\tt BP3} for the process $\mu^- \mu^+ \to \gamma\, \mET$ at the muon collider with $\sqrt{s}=3$ TeV and integrated lumunosity $\mathcal{L}=100$ fb$^{-1}$.}
	\label{tab:cutflow1}
\end{table}
%%%%%
over the large SM background. 
In Fig.~\ref{Fig-3TeV}, we also show the combined SM+BSM normalized distributions of photon energy. For {\tt BP1} and {\tt BP2} with $M_{Z'}=1,2$ TeV, the peak is visible above the falling background distribution while for {\tt BP3} with $M_{Z'}=2.9$ TeV the  peak lies in the softer $E_\gamma$ region where the SM background is very large. 
%%
%%%
\begin{table}[h!]
\centering
\resizebox{8.4cm}{!}{
\begin{tabular}{|c|c|c|c|c|c|c|c|c|}
\cline{2-9}
\multicolumn{1}{c|}{$\mathcal{L}=300$ fb$^{-1}$} & \multicolumn{8}{c|}{$\gamma+\mET$} \\  \hline
                          & \multicolumn{2}{c}{} &  \multicolumn{1}{c}{}&\multicolumn{1}{c}{Events} &  \multicolumn{2}{c}{} &  \multicolumn{2}{c|}{}\\ \hline
                       Cuts & SM & \tt{BP2} & SM & \tt{BP4} & SM &\tt{BP5} & SM & \tt{BP6} \\ \hline
        $N_\gamma=1$ & $565055$  & $75$ & $565055$ & $87$ & $565055$ & $267$ & $565055$ & $2400$\\ \hline
$E_{\gamma} \in E_{\rm{peak}}\mp 200$ GeV & $411$ & $70$ & $591$ & $74$ & $3134$ &  $170$& $466920$ & $2285$\\  \hline 
$E_{\gamma} \in E_{\rm{peak}}\mp 100$ GeV & $156$ & $66$ & $312$ & $65$ & $1581$ &  $130$& $429180$ & $2111$\\  \hline  \hline
    \multicolumn{1}{|c|}{Significance ($\mathcal{S}$)} & \multicolumn{2}{c|}{$4.94$} & \multicolumn{2}{c|}{3.57} &  \multicolumn{2}{c|}{3.24} &  \multicolumn{2}{c|}{3.22} \\ \hline
	\end{tabular}}
 \caption{\small The cutflow information of SM background, {\tt BP2}, {\tt BP4} , {\tt BP5} and 
 {\tt BP6} for the process $\mu^- \mu^+ \to \gamma\, \mET$ at the muon collider with $\sqrt{s}=10$ TeV and integrated lumunosity $\mathcal{L}=300$ fb$^{-1}$.}
	\label{tab:cutflow2}
\end{table}
Therefore, the peak is hardly visible to the naked eye. We however note that for the $Z'$ mass close to $\sqrt{s}$ of the machine, the production of the $Z'$ is practically at resonance and we will get a huge enhancement of the production cross section, as seen in Fig.~\ref{Fig-crx} and Table~\ref{tab:crx}. We therefore note that for a very heavy $Z'$ the background events in the $\gamma+\mET$ channel becomes large, but the signal cross section also gets a significant boost.
In our final analysis of this channel, we impose additional conditions that the events contain only one single hard photon whose energy is constrained to the bins around the peak, as illustrated in Tables~\ref{tab:cutflow1} 
and \ref{tab:cutflow2}.
The effect of the cuts on signal and SM background events are shown in Tables~\ref{tab:cutflow1} and \ref{tab:cutflow2}. 
The signal significance $(\mathcal{S})$ to the SM background has been calculated using the formula \cite{Cowan:2010js},
\begin{equation}
\mathcal{S} = \sqrt{2\left[\left(s+b\right)\ln\left(\frac{s+b}{b}\right)-s\right]},
\end{equation}
where $s$ and $b$ represent the number of signal and background events respectively. The signal significance of the benchmark points has been shown in the last row of Table \ref{tab:cutflow1} \& \ref{tab:cutflow2} for $3$ and $10$ TeV muon collider.
%%%%%%%%%%%%%%%%%%%%%%%%%%%%%%%%%%%%%%%%%%%%%%%%%%%%%%%%%%%%%%%%%%%%%%%%%%%%%%%%%%%%%%%%%
\begin{figure}[b!]
\begin{center}
\includegraphics[width=6cm,height=5cm]{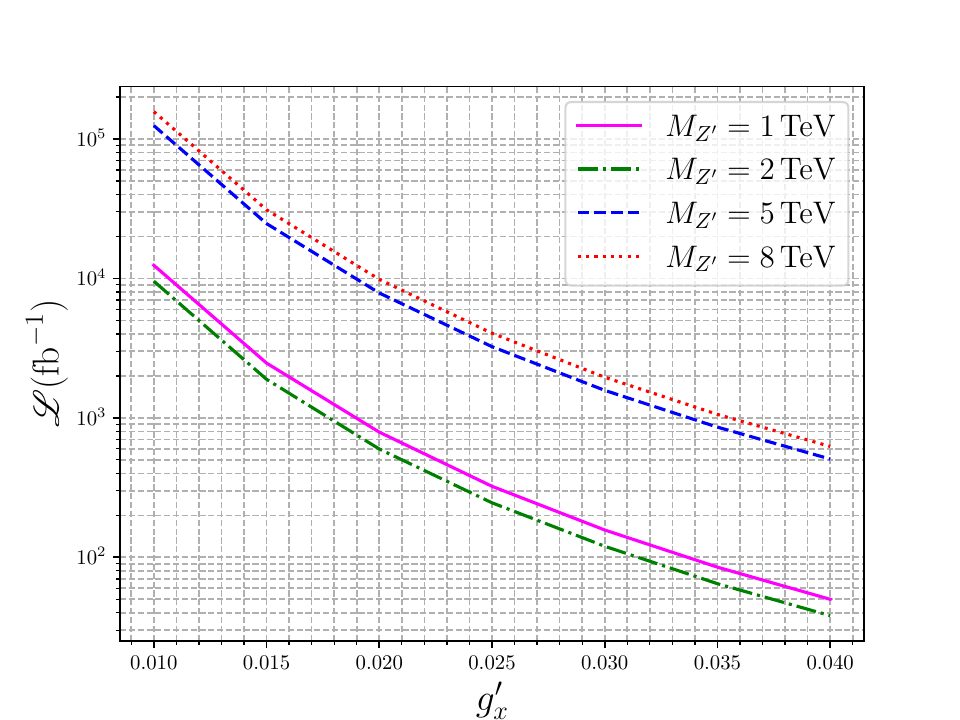}
\end{center}
\vspace{-0.5cm}
\caption{The 3-sigma significance curves of the signal for $M_{Z'} = 1$ and $2$ TeV at $\sqrt{s} = 3$ TeV, and for $M_{Z'} = 5$ and $8$ TeV at $\sqrt{s} = 10$ TeV, are shown as function of the GKM parameter $(g'_{x})$ and luminosity $(\mathcal{L})$. }
\label{Fig-3sigma}
\end{figure}
%%%%%%%%%%%%%%%%%%%%%%%%%%%%%%%%%%%%%%%%%%%%%%%%%%%%%%%%%%%%%%%%%%%%%%%%%%%%%%%%%%%%%%%%%
We find that as long as the $Z'$ mass is below $\sqrt{s}$ of the machine, one will be able to observe the signal of the heavy invisible exotic as well as determine its mass in this search channel at the muon collider.
This can be seen from the values of $\mathcal{S}$ in Tables~\ref{tab:cutflow1} and \ref{tab:cutflow2}. The value for the integrated luminosities are nominal in comparison to what will be available at the proposed muon colliders. Therefore it is quite impressive that how an invisibly decaying exotic with very small couplings to SM particles can be effectively discovered as long as its mass is less than the machine center of mass energy. To show how the signal observation will be sensitive to the $Z'$ mass and its coupling to SM fermions, we plot the $3\sigma$ sensitivity of the signal as a function of the integrated luminosity in Fig.~\ref{Fig-3sigma}. With an integrated luminosity of 
$1~{\rm ab^{-1}}$, the muon collider with $\sqrt{s}=3$ TeV will be able to observe 
a $Z'$ with mass of $2$ TeV with SM couplings as small as $g_x' \simeq 0.018$. This reach will be more as we go to higher mass of $Z'$, where the production cross section rises (Fig.~\ref{Fig-crx}). Similarly, with $\sqrt{s}=10$ TeV a $Z'$ with mass of $8$ TeV can be probed with similar coupling strength. Note that such heavy $Z'$ with the above mentioned coupling strengths will be practically impossible to observe even at the very high luminosity LHC. This search can be extended to the higher $\sqrt{s}$ options of the muon collider to probe even heavier $Z'$.

To conclude, we show how an invisibly decaying heavy BSM exotic can be effectively discovered at the muon colliders in the $\gamma + \mET$ channel using the {\it radiative return} feature that appears in the photon energy distribution. To highlight our result we consider a model which is a $U(1)$ extension of the SM. The new force carrier $Z'$ couples to the BSM particles via the new gauge interaction ($g_x$) while it couples to SM fields via the GKM ($g_x'$) and the $Z-Z'$ mixing angle is assumed to be very small. This model generates small neutrino masses and can accommodate DM. We work on a scenario where the $Z'$ mass is very heavy and $g_x \gtrsim 10 g_x'$ and the $Z'$ decays mostly to the DM and a long-lived particle, therefore becoming invisible. Such a $Z'$ will be difficult to produce and observe at the LHC. We show that its associated production with the photon at the future muon collider at $\sqrt{s}=3, 10$ TeV will be able to determine its mass and distinguish the signal over the SM background by simply looking at the photon energy distribution. The search sensitivity to the $Z'$ mass will depend on the values of both $g_x$ and $g_x'$ but the mass reach is only constrained by the center of mass energy of the muon collider.

\subsection*{Acknowledgments}
SKR would like to acknowledge S. Raychaudhuri for introducing him to the idea of "radiative return" to new physics back in 2000. AS would like to thank HRI for support through the "Visiting Students Program". AKB thanks T. Samui for the technical help. The authors would like to acknowledge support from the Department of Atomic Energy, Government of India, for the Regional Centre for Accelerator-based Particle Physics (RECAPP).

%\bibliography{nuphilZ}
\end{document}